\documentclass[11pt]{article}%
\usepackage{amssymb}
\usepackage{amsmath}
\usepackage{amsthm}
\usepackage{amsfonts}
\usepackage{amsmath, amsthm, amssymb}
\usepackage{parskip}

\DeclareSymbolFont{AMSb}{U}{msb}{m}{n}
\DeclareMathSymbol{\Z}{\mathbin}{AMSb}{"5A}
\DeclareMathSymbol{\R}{\mathbin}{AMSb}{"52}
\DeclareMathSymbol{\C}{\mathbin}{AMSb}{"43}

\newcommand{\E}{\ensuremath{\vec{E}}\  }
\newcommand{\vs}{\ensuremath{\vec{S}}\  }
\newcommand{\B}{\ensuremath{\vec{B}}\  }
\newcommand{\G}{\ensuremath{G_{2, 4}}\  }
\newcommand{\p}{\ensuremath{P_{ab}}\  }
\newcommand{\s}{\ensuremath{S^2 \times S^2}\  }
\newcommand{\h}{\ensuremath{H_{ab}}\  }
\newcommand{\g}{\ensuremath{\tilde{G}_{2,4}}\  }
\newcommand{\jp}{\ensuremath{J_p}}
\newcommand{\sv}{\ensuremath{\vec{S}}\  }
\newcommand{\xv}{\ensuremath{\vec{\xi}^+}\  }
\newcommand{\uv}{\ensuremath{\vec{\xi}^-}\  }
\newcommand{\x}{\ensuremath{\vec{\xi}}\  }
\newcommand{\rh}{\ensuremath{\hat{r}\  }}
\newcommand{\del}{\ensuremath{\partial}}

\newcommand{\uc}{\ensuremath{U_C (1)}\  }

\input epsf.tex
\begin{document}
\bigskip\begin{titlepage}
\begin{flushright}
UUITP-03/07\\
hep-th/yymmnnn
\end{flushright}
\vspace{1cm}
\begin{center}
{\Large\bf The Grassmannian Sigma Model in \\  $SU(2)$\ Yang-Mills Theory\\}
%{\Large En bra undertitel\\}
\end{center}
\vspace{3mm}
\begin{center}
{\large
David Marsh{$^*$}} \\
\vspace{5mm}
%{\large
%David Marsh\footnote[*]{dama3517@student.uu.se} } \\
%\vspace{5mm}
\emph{Department of Theoretical Physics, Uppsala University} \\
\emph{SE-751 08, Uppsala, Sweden} \\
\vspace{5mm}
\end{center}
\vspace{5mm}
\begin{center}
{\large \bf Abstract}
\end{center}
\noindent
Spin-charge separation in pure   $SU(2)$\ Yang-Mills theory was recently found to involve the dynamics of an  $O(3)$\ non-linear sigma model and, seemingly, a Grassmannian non-linear sigma model. In this article we explicitly construct the Grassmannian sigma model of the form appearing in the the spin-charge separated  $SU(2)$\ theory through a quaternionic decomposition of the manifold, thus verifying its relevance in this context. The coupling between this model and the  $O(3)$\ non-linear sigma model is further commented upon. \vfill
\begin{flushleft}

\tt
$^*$dama3517@student.uu.se
\end{flushleft}
\end{titlepage}\newpage

\section{Introduction}
In a recent article by Faddeev and Niemi \cite{FN}, spin-charge separation in pure   $SU(2)$\ Yang-Mills theory was investigated. Through a non-linear change of variables, the gluon field was split into two charged scalar fields on the one hand, and an uncharged spin--carrying field on the other. Similar approaches are well-known from condensed matter physics, where the charged excitations are called \emph{holons}, while the spin--carrying particles are called \emph{spinons}. It was found that the Yang-Mills theory counterparts of the spinon and the holon were subject to a mutually attractive compact $U(1)$\  force, which is known to become strong at short distances. This provides a rationale for the physical relevance of the phenomenon.  It is reasonable to expect that in the  confining phase the gluon would seem pointlike, while under very particular circumstances, involving  the presence of a non-vanishing holon condensate, the gluon might be split.

The possibility of this condensate  holding the clues for understanding confinement in the low-energy regime of the theory was discussed in \cite{FN}, \cite{niemi}. Its prospects for being the missing mechanism for explaining high-temperature superconductivity is discussed in \cite{niemi},\cite{Andersson}. Its role as the scale factor of a conformally flat space-time and its implications for theories of gravity was discussed in \cite{FN}, \cite{sergei}. 

Interpretations apart, the dynamics of the splitted gluons, i.e. the pure  $SU(2)$\ Yang-Mills Lagrangian in spin-charge separated variables, contains a form of the  $O(3)$\ non-linear sigma model, coupled to the real Grassmannian non-linear sigma model. The Grassmannian present is the \G, the manifold of two-dimensional subspaces of a four dimensional vector space. This latter manifold is important in the interpretation of the internal $U(1)$\  force, since it gives significant contributions to the field strength and seems on the whole to be a key to understanding spin-charge separation in gauge theories.

Yet the form of the Grassmannian sigma model appearing in \cite{FN} is surprising, and, to our knowledge, is not previously discussed in the litterature. It is therefore of value to show by geometric construction that the Grassmannian sigma model may indeed be written on the form suggested by Faddeev and Niemi, as we do in this note.

 We also deepen  the study of the coupling between this model and the  $O(3)$\ non-linear sigma model, which in \cite{FN} only was interpreted in certain limiting cases. As will be discussed, the interaction terms may in general be written as a function of local sections of the holomorphic cotangent bundle of the spheres factorizing the oriented Grassmannian.

\section{Spin-Charge Decomposition}     
Following \cite{FN}, we will be considering  $SU(2)$\ Yang-Mills theory over Euclidean space-time $(\R^4, \delta)$\ with indices $a, b$.

The  $SU(2)$\ Yang-Mills gauge field $A$\   can be written in terms of the linear combinations 
\[
\mathcal{A}_a = A^i_a \frac{\sigma^i}{2} = A_a \frac{\sigma^3}{2}  + X^+_a \frac{\sigma^-}{2}  + X^-_a \frac{\sigma^+}{2}. 
\]  
The $\sigma^i$\  are the Pauli matrices while the fields and off-diagonal projectors are denoted
\begin{align*}
A_a &= A^3_a \\
 X^{\pm}_a &= A^1_a \pm i A^2_a  \\
\sigma^{\pm} &= \frac{1}{2}(\sigma^1 \pm i \sigma^2).
\end{align*}
Under an infinitesimal gauge transformation $U = e^{i \theta_a \sigma^a /2}$, with the choice of covariant derivative $D^+ = \del + i A$, the gauge fields transform as
\begin{align}
\delta_{\theta} A^i_a =  \del_a \theta^i + \epsilon^{ijk} A^j_a \theta^k \label{gtf1}.
\end{align}
Consider gauge transformations along the $\sigma^3$\   direction only. These transformations form an $U(1)$\  subgroup of $SU(2)$, denoted \uc, where $C$\ stands for colour. For an infinitesimal $h \in \uc$,
\[
h = e^{i \omega \sigma^3 / 2} \label{uc}.
\]
According to (\ref{gtf1}), the gauge fields now transform as
\begin{align}
\delta_h A_a &= \del_a \omega \notag \\
\del_h X^{\pm} &= \mp i \omega X^{\pm}_a \label{uc1}.
\end{align}
So $A^a$\ transforms as the \uc gauge field and the off-diagonal components as \uc charged fields. The third component of the $SU(2)$\ field strength is
\begin{align}
F^3_{ab} = & \partial_a A^3_b - \partial_b A^3_a + \epsilon^{3jk} A^j_a A^k_b = \notag \\ 
 = & \partial_a A_b - \partial_b A_a + \frac{i}{2} ( X^+_a X^-_b - X^+_b X^- _a) = F_{ab} + P_{ab},
\end{align}
 where $F_{ab}$\   is the $U(1)$\  field strength and $P_{ab}$\   is given by
  \begin{align}
P_{ab} =   \frac{i}{2} ( X^+_a X^-_b - X^+_b X^- _a)  = A^1_a A^2_b - A^1_b A^2_a \label{P}.
\end{align}
This second rank tensor has a natural interpretation in terms of the Grassmannian \G of unoriented two dimensional planes through the origin of a four dimensional vector space. Since $SO(4)$ (with the standard action) acts transitively on  $G_{2,4}$, and the stabilizer at $p \in \G$ is given by rotations in the plane $p$\ as well as in the plane orthogonal to $p$, the representation of the manifold as a homogeneous space is 
\[
\G \simeq \frac{SO(4)}{SO(2) \times SO(2)}.
\]
Analogous to the homogeneous coordinates of real projective space, $p \in \G$ can also be represented by a $2 \times 4$ \ matrix $M$\ whose row-vectors span $p$. Let $A^1,\ A^2 \in \R^4$\ span $p$. A matrix representative of $p$\ is given by
\[
M = \left(
\begin{matrix}
A^1_1 & A^1_2 & A^1_3 & A^1_4 \\
A^2_1 & A^2_2 & A^2_3 & A^2_4 
\end{matrix} \right).
\]
The representation is not unique since, if equivalence is taken to mean representing the same element of the Grassmannian, then for all $ g \in GL(2, \R)$
\begin{align}
M \sim g \ M.  \label{ekviv} 
\end{align}
Thus, the manifold can be represented as the coset space
\begin{align*}
 \G \simeq \frac{M_{2,4}}{GL(2, \R)},
\end{align*}
where $M_{2, 4}$\ denotes the set of $2 \times 4$\ matrices of maximal rank.

Standard local coordinates (equivalents of the inhomogeneous coordinates for the projective space) are obtained in six charts for the pairs $(i, j)\  \  i,j = 1, \ldots, 4$. Since $M$\ is of maximal rank, at least one of its minors is non-vanishing, and, equivalently, at least one of its six $2 \times 2$\  column matrices, $m_{ij}$, is invertible. To be explicit, consider the chart $(1, 2)$, and suppose that the inverse of the matrix formed out of the first and second column of $M$\ exists. Denote it $m_{12}^{-1}$.
\begin{align*}
M' = m_{12}^{-1}\   M = \left(
\begin{matrix}
1 & 0 & \frac{P_{32}}{P_{12}} & \frac{P_{42}}{P_{12}} \\
0 & 1 & \frac{P_{13}}{P_{12}} & \frac{P_{14}}{P_{12}} 
\end{matrix} \right),
\end{align*}
where equation (\ref{P}) has been used. It follows that 
\begin{align}
P_{12} P_{34} + P_{13} P_{42} + P_{14} P_{23} = 0\ .
\end{align}
This is the Pl\"ucker equation. It describes a quadratic hyper-surface in $\R^6$\  called the Klein Quadric. The anti-symmetric tensor \p with six independent entries is called the Pl\"ucker Compound  and its entries the Pl\"ucker coordinates. The equivalence relation, (\ref{ekviv}), turns into
\begin{align*}
 P_{ab} \sim \det g \ P_{ab}. 
\end{align*}
So, supplying the Grassmannian with Pl\"ucker coordinates embeds the manifold as the Klein Quadric in $\R P^5$. 

Define the complexified four-component vector $v_a = A^1_a + i A^2_a$. Note that equation (\ref{ekviv}) implies that
\begin{align}
 v_a \sim w\ v_a,\  \    \   \   w \in \C \label{cp}.
\end{align}
That is,  $v_a$\ is an element of the three dimensional complex projective space, $\C P^3$. Represent $p \in \G$ by an orthonormal frame, $e^1_a,\   e^2_a$\ and introduce the spinon field $e_a = \frac{1}{\sqrt{2}}(e^1_a + i e^2_a)$\   that satisfies
\begin{align}
e_a \ e_a = 0 \\
\bar{e}_a\ e_a = 1.
\end{align}
Note that by these restrictions $w$\ in (\ref{cp}) turns into a phase. 

%%%%%%%%%%%%%%%%%%%%%%
Spin-charge separation\footnote{Using a different separation between the spin and the charge variables, it has been found that $SU(2)$\  Yang-Mills theory possesses features similar to those of liquid crystals, see \cite{chernodub}} is obtained through the change of variables
\begin{align}
X^+_a =&  \psi_1 e_a + \psi_2 \bar{e}_a \label{explus} ,
%X^-_a =& \psi_2^* e_a + \psi_1^* \bar{e}_a
\end{align} 
where $\psi_{1, 2} $ are complex valued scalars called holons. We rewrite the Pl\"ucker compound in these coordinates to obtain
\begin{align}
P_{ab} = \frac{i}{2} ( |\psi_1|^2 - |\psi_2|^2 ) ( e_a \bar{e}_b - e_b \bar{e}_a )= \notag \\ =
\rho^2 \cos \theta \  \frac{i}{2} (e_a \bar{e}_b - e_b \bar{e}_a) = \rho^2 t_3 H_{ab} \label{H},
\end{align} 
where the condensate $\rho$\ and the real vector $\vec{t}$ \ are defined as
\begin{align}
\rho^2 =& |\psi_1|^2 + |\psi_2|^2  \\
\vec{t} =& \frac{1}{\rho^2}\  \left( \psi^*_1 , \psi^*_2 \right)\  \vec{\sigma} \  \left( \begin{array}{c} \psi_1 \\ \psi_2 \end{array} \right) = \left( \begin{array}{c}
\cos \phi \sin \theta \\ 
\sin \phi \sin \theta \\
\cos \theta
\end{array}
\right). 
\end{align}
The factorization of the Pl\"ucker compound into a pre-factor and the tensor $H_{ab}$\ corresponds to restricting the representation of the Grassmannian to orthonormal pairs of spanning vectors. The tensor \h is invariant under the changes of matrix representative, $M$, in (\ref{ekviv}). 

In analogy with the electromagnetic field strength, the ''electric'' and ''magnetic'' fields are defined as
\begin{align}
(\E)_i =& H_{0i} \\
(\B)_i =& \frac{1}{2} \epsilon_{ijk} H_{jk}\   \   ,
\end{align}
satisfying
\begin{align}
\E \cdot \B = 0 \label{klein} \\
\E^2 + \B^2 = 1 \label{norm} \\
(\E, \B) \sim -(\E, \B) \label{antipod}.
\end{align}
In this notation, (\ref{klein}) is the Pl\"ucker equation, while the  constraints (\ref{norm}) and (\ref{antipod}) reduce the space to $\R P^5$\ by identifying anti-podal points on $S^5$. Thus we can write
\begin{align}
\begin{cases}
\E &= \cos \vartheta \  \hat{k} \label{vartheta} \\
\B &= \sin \vartheta \   \hat{l} \\
\vs &= \E \times \B .
\end{cases}
\end{align}
For future reference we will also introduce a parametrization of the holons
\begin{align}
\psi_1 = \rho e^{i \xi} \cos \frac{\theta}{2} e^{-i \phi / 2} \label{sph} \\
\psi_2 = \rho e^{i \xi} \sin \frac{\theta}{2} e^{i \phi / 2} .
\end{align}

Although \p is invariant, the spinon $e_a$\  picks up a phase under rotation of the orthonormal frame $e^1, e^2$\  in (\ref{cp}), so the off-diagonal gauge fields are invariant under $U_I (1) \subset SL(2, \R)$ \ acting as
 \begin{align}
 A_{a} \rightarrow A_{a},\   \  & \   X^+ \rightarrow X^+ \notag  \\
 \psi_1 \rightarrow& e^{i \lambda} \psi_1 \label{uep1} \\
 \psi_2 \rightarrow& e^{-i \lambda} \psi_2 \label{uep2} \\
 %\phi \rightarrow& \phi - 2 \lambda \\
 %t_3 \rightarrow& t_3 \\
 e \rightarrow& e^{-i \lambda}  e \label{uee} .
 %t_{\pm} := t_1 \pm i t_2 \rightarrow& e^{\mp i \lambda} t_{\pm} \label{tf2}
 \end{align} 
 Clearly $\rho^2$\ and $\vec{t}$ \ are invariant. A suitable  $U_I(1)$\  gauge field is
 \begin{align}
 C_a = i \bar{e}_b \del_a e_b \notag \\
 C_a \rightarrow C_a + \del_a \lambda .
 \end{align}
 Also note that
 \begin{align}
 t_{\pm} := t_1 \pm i t_2 \rightarrow& e^{\mp i \lambda} t_{\pm} \label{tf2}.
 \end{align}
 This motivates the introduction of an $\uc \times U_I(1)$\ invariant vector $\vec{n}$. Take 
 \begin{align}
 e_0 = e^{i \eta} |e_0|  \label{eta} \\
 e_a = e^{i \eta}\   \hat{e}_a ,
 \end{align}
 so that $\eta \rightarrow \eta - \lambda$\ in (\ref{uee}). Then $\vec{n}$, which will describe the $O(3)$\  non-linear sigma model, can be defined as
 \begin{align*}
 n_{\pm} =& e^{2 i \eta} t_{\pm} \\
 n_{3} =& t_{3} .
 \end{align*}
 
In terms of these new coordinates, the \uc transformation, (\ref{uc}) is
 \begin{align}
 A_{a} \rightarrow A_a + \del_a \omega,&\   \   X^{\pm} \rightarrow e^{\mp i \omega } X^{\pm}
  \label{ag} \\
 \psi_1 \rightarrow& e^{- i \omega} \psi_1 \label{ucp1}\\
 \psi_2 \rightarrow& e^{- i \omega} \psi_2 \label{ucp2}\\
 e_a \rightarrow& e_a \label{uce} .
 \end{align}
 For \uc the gluon field degrees of freedom are described by the multiplet
 \begin{align*}
 A^i \sim ( A_a, \psi_1 , \psi_2 , e_a ) .
 \end{align*}
The first field is the gauge field, while the second two are charged but spin-less complex fields that transform with the same charge under \uc. These holons are oppositely charged with respect to the internal force, giving rise to a mutual attraction between the holons and the spinon, in the decomposition (\ref{explus}). The spinons on the other hand do not carry colour charge but transform projectively under Lorentz ($SO(4)$) transformations \cite{FN}.

Considering the $U_I (1)$ transformations the multiplet structure is
  \begin{align*}
  A^i \sim ( C_a, \psi_1 , \psi_2, A_a). 
 \end{align*}

The dynamics of the gluon fields in this decomposition are stipulated by the classical  $SU(2)$\ Yang-Mills Lagrangian with the off-diagonal gauge fields fixed (Maximum Abelian Gauge)
 \begin{align}
\mathcal{L}_{YM} = \frac{1}{4} (F^i_{ab})^2 + \frac{1}{2} | D^+_{a} X^+_a|^2 .
\end{align}
Re-expressing it in spin-charge separated, $\uc \times U_I(1)$\ invariant variables leads to \cite{FN}
\begin{align}
\mathcal{L}_{YM} = \frac{1}{4} \mathcal{F}^2_{ab} + \frac{1}{2} \rho^2 J^2_a + \frac{1}{8} \rho^2 (D^{\hat{C}}_a \vec{n} )^2 + \frac{\rho^2}{4} \{ (\del \E)^2 + (\del \B)^2 \} \notag \\
+ \frac{1}{4} \rho^2 \{ n_+ (\del_a  \hat{\bar{e_b}})^2 + n_- (\del_a \hat{e}_b )^2 \} + \frac{3}{8} (1 - n_3 )\rho^4 - \frac{3}{8} \rho^4 \label{L},
\end{align}
where
\begin{align}
\mathcal{F}_{ab} &= \del_a J_b  - \del_b J_a + \frac{1}{2} \vec{n} \cdot \del_a \vec{n} \times \del_b \vec{n} - \{ \del_a (n_3 \hat{C}_a - \del_b (n_3 \hat{C}_a) \} - 2 \rho^2 n_3 H_{ab} \\
J_a &= \frac{i}{2 \rho^2} \{ \psi^*_1 D_a \psi_1 - \psi_1 \bar{D}_a \psi^*_1 + \psi^*_2  D_a \psi_2 - \psi_2 \bar{D}_a \psi^*_2\}.
\end{align}
%%%%%%%%
The covariant derivatives and the connection are defined as
\begin{align}
D^{\pm}_a = \del_a \pm i A_a \\
D^C_a \psi_1 = ( \del_a + i A_a - i C_a) \psi_1 \\
D^C_a \psi_2 = ( \del_a + i A_a + i C_a) \psi_2 \\
D^C_a e_b = ( \del_a +  i C_a) e_b \\
\hat{C}_a = i \hat{\bar{e}} \cdot  \del_a \hat{e}_a = C_a + \del_a \eta.
\end{align}

The terms of present interest include
\begin{align}
 \frac{\rho^2}{4} \left( (\del \E)^2 + (\del \B)^2 \right) \label{grassman},
\end{align}
which is the ostensible \G Grassmannian sigma model in \cite{FN}.
The coupling between this model and the $O(3)$\ non-linear sigma model described by $\vec{n}$\  is given by
\begin{align}
\frac{\rho^2}{4 } \{n_+ ( \del_a \hat{e}^*_b )^2 + n_- (\del_a \hat{e}_b)^2 \} . \label{int}
 \end{align}

\section{Quaternionic Decomposition}
If (\ref{grassman}) actually corresponds to the \G sigma model, the metric tensor of the six-dimensional space of Pl\"ucker coordinates (\E and \B space) should  be taken to be flat, at least in the neighbourhood of the embedded Grassmannian. Consistently joining this fact with $(A^1_a, A^2_a) \in (\R^4 \times \R^4, \delta)$\ may formally be stated as finding a solution to a set of 15 coupled partial differential equations. We will not pursue this direct approach but instead induce the metric tensor on  \G by using the spherical factorization of the  Grassmannian of \emph{oriented}  two dimensional subspaces of a four dimensional vector space, \g. The oriented Grassmannian is the two-fold covering of \G and satisfies equations (\ref{klein}), (\ref{norm}), but not (\ref{antipod}). It is embedded as the Klein Quadric in $S^5$.

There are several well-known ways of doing this decomposition, the arguably simplest is by representing each $p \in$\g by a decomposable element of $\Lambda^2(\R^4)$. The Hodge star dual operator maps each oriented plane to its oriented orthogonal complement. Expressing $p$\ in the eigenbasis of the Hodge star exhibits the product sphere nature of \g.

In this context however, it is fruitful to take $p \in$ \g spanned by orthonormal basis $(e^1, e^2)$\  and identify these vectors with the \emph{quaternions} $(H, \delta)$\  in an obvious way 
\begin{align*}
e_a = e_0 +i e_1+j e_3+k e_3 = e_0 + \vec{e},
\end{align*}  
with $i^2 =j^2=k^2=ijk= -1$. The quaternionic multiplication rules may be expressed as, for $p, q \in (H, \delta)$
\begin{align}
p &= a + \vec{u} \   \   \   \    q = b + \vec{v} \notag \\
pq &= ab -\vec{u} \cdot \vec{v} + a\vec{v} + b \vec{u} + \vec{u} \times \vec{v} \label{qmult} .
\end{align}
Two independent divisions of the spanning quaternions can be made
\begin{align}
\left( \frac{e^2}{e^1} \right)_R =  e^2 \bar{e}^1 = \E - \B = \uv \label{divr} \\
\left( \frac{e^2}{e^1} \right)_L =  \bar{e}^1 e^2 = \E + \B = \xv \label{divl} .
\end{align}
Since \x$^{\pm} = 1$, these are maps from \g to $\s \subset (H \times H, \delta)$. The spheres are standardly interpreted as being the sphere of complex structure of $\R^4$\ and a trivial $\C P^1$\ bundle.

To see this, associate with $p$\ an almost complex structure consisting of positive $\frac{\pi}{2}$\ rotations of the vectors in $p$\ and its orthogonal complement, and extend this rotation linearly to the entire space $H$. Denote this almost complex structure $J_p$. Then $(J_p)^2 = -1$.

Under the action of \jp, the real unit quaternion $u = 1 + \vec{0}$ will be rotated into some unit quaternion orthogonal to it. 
\begin{align*}
\jp(u) = v(p) \in S^2,
\end{align*}
where the sphere is the sphere of imaginary unit quaternions. The pair $(u, v(p))$\ can be used to label the complex structures on $\R^4$, and the sphere is the set of complex structure of $\R^4$. Since $\jp: (e^1, e^2) \rightarrow (e^2, -e^1)$, we can identify $v(p) \in S^2$ explicitly
\begin{align}
v(p) = \left( \frac{e^2}{e^1} \right)_R =  e^2 \bar{e}^1 = \E - \B = \uv .
\end{align}

The elements $\tilde{p}$\ of the Grassmannian sharing  complex structure with $p$\  satisfy
\begin{align*}
V \in \tilde{p}\   \   \jp (V) \in \tilde{p}.
\end{align*} 
In other words, the set of $\tilde{p}$\ sharing a complex structure is the set of one-dimensional complex subspaces with respect to \jp, i.e. $\C P^1 \simeq S^2$. The independent coordinate on this sphere is given by (\ref{divl}).

In conclusion we can write $(\xv, \uv) \in \s \subset (H \times H, \delta)$\ and induce the Euclidean metric on the spheres to get the standard round metric in spherical local coordinates $(\alpha_1, \beta_1, \alpha_2, \beta_2)$
\begin{align*}
ds^2 = d\alpha_1^2 + \sin^2 \alpha_1 \ d\beta_1^2 + d\alpha_2^2 + \sin^2 \alpha_2 \  d \beta_2^2 .
\end{align*}
Consistently we can take the metric of the six-dimensional space of independent Pl\"ucker coordinates to be flat, at least in the neighbourhood of the Grassmannian. This results in the non-linear sigma model Lagrangian
	\begin{align}
	\mathcal{L}_{\G} = g_{ab}\   \del_a p^i \ \del_b p^i = ( \partial \E)^2 + ( \partial \B )^2 ,
	\end{align}
	with $p^i,\  i = 1, \ldots, 6$\ being the Pl\"ucker coordinates in agreement with \cite{FN}. The subsidiary constraints are given by (\ref{klein})-(\ref{antipod}), such that \G $\simeq \R P^2 \times \R P^2$.

\section{The Interaction}
The  $O(3)$\ non-linear sigma model appearing in the decomposition is given by
	\begin{align*}
	(\del \vec{n} )^2 + ( \vec{n} \cdot (\del_a \vec{n} \times \del_b \vec{n} ))^2 ,
	\end{align*}
for the unit vector $\vec{n}$. This model has since long been thought to be related to the low-energy regime of SU(2) Yang-Mills, and it is known to support stable knotted solitons, as discussed in [1]. The coupling between the  $O(3)$\ non-linear sigma model and the Grassmannian model is quite intricate, and was only analysed in \cite{FN} in the purely electric ($\vartheta \rightarrow 0$ in (\ref{vartheta}))  and purely magnetic ($\vartheta \rightarrow \pi/2$)  limits. We will analyse the situation in general and will be able to assert that the coupling is related to local sections of the holomorphic cotangent bundle ($T^{*(1, 0)}S^2$) of the spheres factorizing the Grassmannian.  

 In the two sets of spherical coordinates $( \alpha_1, \beta_1, \alpha_2, \beta_2)$\ of the two spheres of \g, these forms are given by
\begin{align}
dz_i =& \ d \alpha_i + i \sin \alpha_i \  d \beta_i ,\   \    \   i = 1, 2\label{f+} .
\end{align}

The coupling (\ref{L}), can be elaborated by plugging in the expression for $e$\ in terms of the electric and magnetic fields,
\begin{align}
\frac{ \rho^2 n_{+}}{128 \sv^2}  \left( \partial_a ( \E + \B)\right) \cdot \left( \E- \B - 2i \sv \right)\left( \partial_a (\E - \B)\right) \cdot \left( \E + \B - 2i \sv\right)  + \notag \\
+ \frac{ \rho^2 n_{-}}{128 \sv^2} \left( \partial_a ( \E + \B)\right) \cdot \left( \E- \B + 2i \sv \right)\left( \partial_a (\E - \B)\right) \cdot \left( \E + \B + 2i \sv \right) .
\end{align}
Expressed in the spherical unit vectors,
\begin{align}
\frac{ \rho^2 n_{+}}{128 (\sin \vartheta \ \cos \vartheta)^2} \left( \partial_a (\xv)\right) \cdot \left( \uv + i (\xv \times \uv) \right)\left( \partial_a (\uv)\right) \cdot \left( \xv + i(\xv \times \uv)\right)  + \notag \\
+ \frac{ \rho^2 n_{-}}{128 (\sin \vartheta \ \cos \vartheta)^2} \left( \partial_a ( \xv)\right) \cdot \left( \uv - i (\xv \times \uv) \right)\left( \partial_a (\uv)\right) \cdot \left( \xv - i (\xv \times \uv)\right) \label{56b} .
\end{align}
For the moment, identify 
\begin{align*}
\xv = \hat{r}_1 = \left( \begin{array}{c}
\cos \beta_1 \sin \alpha_1 \\
\sin \beta_1 \sin \alpha_1 \\
\cos \alpha_1
\end{array}
\right),
\end{align*}
where the components are taken with respect to the standard Cartesian basis. Expressing $\uv$ in the orthonormal triplet $(\hat{r}_1, \hat{\beta}_1, \hat{\alpha}_1)$, of the $S^2_+$  gives 
\begin{align}
\uv = \cos 2\vartheta \ \hat{r}_1 + \sin 2\vartheta ( \cos \tau \ \hat{\beta}_1 + \sin \tau \ \hat{\alpha}_1),
\end{align}
where the degrees of freedom previously described by the pair $(\alpha_2, \beta_2)$ now are replaced by the angles $(\vartheta, \tau)$. We have for $a=1, \ldots,4$  
\begin{align}
\partial_a (\rh) dx^a = \partial_i (\rh) d \alpha_1^i = \sin \alpha_1 \ d \beta_1 \  \hat{\beta}_1 - d \alpha_1 \  \hat{\alpha}_1.
\end{align} 
Using that
\begin{align}
\partial_a (\xv) \cdot (\uv + i \xv \times \uv) dx^a= \notag \\
=\partial_i (\hat{r}) d\alpha_1^i  \cdot \left( \sin 2\vartheta ( \cos \tau \hat{\beta_1} + \sin \tau \hat{\alpha_1}) + i ( \rh \times \sin 2 \vartheta ( \cos \tau \hat{\beta_1} + \sin \tau \hat{\alpha_1}) ) \right) = \notag \\
= \sin 2 \vartheta \left( -(\sin \tau + i \cos \tau) d \alpha_1 + \sin \alpha_1 (\cos \tau - i \sin \tau) d \beta_1 \right) = \notag \\
= \sin 2 \vartheta \  e^{-i(\tau + \frac{\pi}{2})} \left( d\alpha_1 + i \sin \alpha_1 d\beta_1 \right) \notag ,
\end{align} 
the interaction terms can be reformulated.  Write the terms with the derivative acting on the $S^2_+$ \  vector in terms of the angles $( \beta_1, \alpha_1, \vartheta, \tau)$\ as above, while the terms with the derivative acting on  the $S^2_-$\  vector are expressed in $(\alpha_2, \beta_2, \vartheta, \tau')$.  The inner product in the cotangent space is Euclidean, so the first term of (\ref{56b})  turns into
\begin{align}
%\frac{1}{32} \rho^2 n_+ \ e^{i(\tau' - \tau)} \ ( \del_a (\alpha_1) + i \sin \alpha_1 \del_a \beta_1) ( \del_a \alpha_2 - i \sin \alpha_2 \del_a \beta_2) = \\
\frac{1}{32} \rho^2 n_+ \  e^{i(\tau'- \tau)} \ (dz_1, d \bar{z}_2) . \label{y1} 
\end{align}
%\begin{align}
%\frac{1}{32} \rho^2 n_- \ e^{-i(\tau- \tau')} \ (d\bar{z}_1, d z_2) \label{y2}
%\end{align}
The full interaction between the Grassmannian sigma model and the  $O(3)$\ sigma model is given by,
\begin{align}
 \frac{1}{16} \ \rho^2 \ \Re \{  n_+  e^{i( \tau' - \tau)} \ (d z_1, d \bar{z}_2) \} =
 \frac{1}{16} \ \rho^2 \ \sin \theta\   \Re \{  e^{i(\phi - 2 \eta + \tau' - \tau)} \ (d z_1, d \bar{z}_2) \} 
\end{align}
where the angles $(\theta, \phi, \eta)$\ correspond to the degrees of freedom of the  $O(3)$\ sigma model, as defined in equations (\ref{sph}) and (\ref{eta}). This establishes how the Grassmannian sigma model of \G, as it appears in the spin-charge separated $SU(2)$\ Yang-Mills Lagrangian, relates to the $O(3)$\ non-linear sigma model.

\section{Discussion}
We have shown that the Grassmannian sigma model \G appears in the spin-charge separated SU(2) Yang-Mills theory, and that the interaction can in general be expressed as involving the inner product of the local sections of the holomorphic and anti-holomorphic cotangent bundles of the constituent spheres of \g. 

In \cite{chernodub}, through a \emph{different} spin-charge separation in a \emph{different} gauge, the Yang-Mills theory was cast in a form very similar to the condensed matter system of nematic liquid crystals. Interestingly, also the current decomposition mimics this property.

The product manifold decomposition of the unoriented Grassmannian, \G $\simeq \ \R P^2 \times \R P^2$, is naturally interpreted as the product of two coupled nematic crystals. Particularly, in the manifestly Lorentz invariant vacuum of the theory ($\theta \rightarrow 0$) \cite{FN}, the crystals decouple and the $O(3)$ non-linear sigma model no longer contributes to the dynamics. Possible topological defects are typically classified by the homotopy of the vacuum manifold, and the existence of non-trivial vortices in nematic crystals in three dimensional space corresponds to the non-trivial element of $\pi_1 (\R P^2) = \Z_2$. This suggests that also the Grassmannian part of the decomposed Yang-Mills theory may support non-trivial vortices. 

 These results are parts of a larger programme on understanding and interpreting the structure emerging from the Yang-Mills Lagrangian in spin-charge separated variables. 

\section{Acknowledgements}
I would like to thank prof. Antti Niemi for presenting me this problem and for discussions. I also owe  thanks to Sergey Slizovskiy for discussions and translations and in particular to Maxim Zabzine for discussions and clarifications. I would like to thank Marit Str\"omberg and Olof Ohlsson-Sax for proof-reading.

\end{document}